\documentclass[twocolumn,letter]{jpsj2}

\title{%
YbRh$_2$Si$_2$: Quantum tricritical behavior in itinerant electron systems}

\author{%
Takahiro \textsc{Misawa}\thanks{E-mail:misawa@solis.t.u-tokyo.ac.jp},
Youhei \textsc{Yamaji}, and Masatoshi \textsc{Imada}  
}

\inst{
 {\it Department of Applied Physics, University of Tokyo,
7-3-1 Hongo, Bunkyo-ku, Tokyo, 113-8656, Japan}
}

\recdate{\today}

\abst{We propose that proximity of the first-order transition manifested by 
the quantum tricritical point (QTCP) explains non-Fermi-liquid 
properties of YbRh$_2$Si$_2$. Here, at the QTCP, a continuous phase 
transition changes into first order at zero temperature. The non-Fermi-liquid
behaviors of YbRh$_2$Si$_2$ are puzzling in two aspects; diverging 
ferromagnetic susceptibility at the antiferromagnetic transition and 
unconventional power-law dependence in thermodynamic quantities. These 
puzzles are solved by an unconventional criticality derived from our 
spin fluctuation theory for the QTCP.
}

\kword{quantum critical phenomena, quantum tricritical point, YbRh$_2$Si$_2$, 
non-Fermi-liquid behavior, self-consistent renormalization theory
}

\begin{document}
\maketitle
\begin{figure}[h!]
	\begin{center}
		\includegraphics[width=7.0cm,clip]{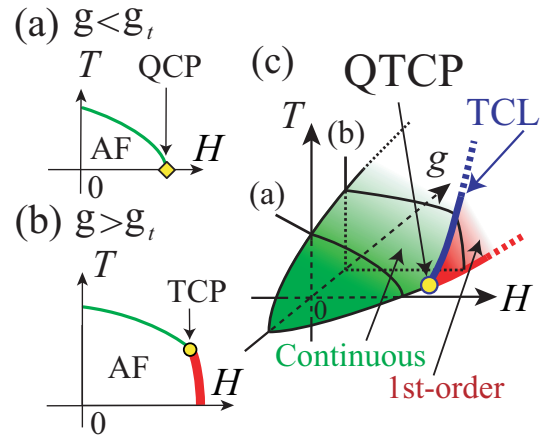}   
	\end{center}
\caption{(color online). (a) Phase diagram of antiferromagnetic (AF) phase with  
critical line [solid (green) curve] ending at the
QCP [(yellow) diamond] in $T$-$H$ plane,
where $T$ ($H$) represents temperature 
(magnetic field).
(b) Phase diagram with TCP [(yellow) circle] 
separating the continuous [thin (green) curve] and first-order [thick (red) curve] transition lines.
(c) Global phase diagram with tricritical line (TCL) separating the surfaces of continuous 
[above TCL (green)] and the first-order [below TCL (red)] surfaces. 
Here $g$ represents parameter to control quantum fluctuations.
In YbRh$_{2}$Si$_{2}$, $g$ may correspond to pressure
measured from the ambient one. 
The QTCP (circle) appears at $(g,H,T)=(g_{t},H_{t},0)$, namely the endpoint of TCL.
Cross sectional view at constant $g$
for $g<g_{t}$[$g>g_{t}$] corresponds to the phase diagram (a)[(b)].
}
\label{fig:QTCP_Global_Phase}
\end{figure}

Critical temperatures of the symmetry-breaking
phase transitions can be
lowered to zero at the quantum critical point (QCP) by tuning quantum fluctuations 
such as by magnetic fields
as shown in Fig. \ref{fig:QTCP_Global_Phase}(a).
Quantum critical phenomena in metals have attracted much interest 
from both theoretical and experimental 
points of view, because of 
not only its own right but also 
unconventional superconductivity as well as non-Fermi-liquid behavior 
observed near the QCP~\cite{Steglich,Stewart}.

The conventional spin fluctuation theory of the 
QCP by Moriya, Hertz and Millis~\cite{Moriya,Hertz,Millis,Stewart} 
has succeeded in explaining a number of non-Fermi-liquid 
properties. However, this picture has been challenged by
many recent experiments~\cite{Steglich,Stewart,Lohneysen},
where criticalities of thermodynamic and transport properties 
do not follow it.

A typical heavy-fermion compound YbRh$_{2}$Si$_{2}$~\cite{Steglich}
belongs to such an unconventional category. 
At the magnetic field $H=0$, it exhibits an antiferromagnetic (AF) transition
at the N${\rm \acute{e}}$el temperature $T_{\rm N}=0.07$K.
An AF 
QCP emerges at
the critical magnetic field $H_c\sim 0.06$T along the $c$ axis~\cite{Trovarelli,Gegenwart_1}.
Near $H_{\rm c}$,
Sommerfeld coefficient of specific heat $\gamma$ is  
logarithmically increased with lowering temperature $T$ above 0.3K and even faster below it~\cite{Custers} 
in contrast to the conventional theory predicting 
convergence to a constant.
Transport and optical data 
roughly show the resistivity linearly 
scaled with $T$ and frequency~\cite{Trovarelli}.
Among all, a key aspect is an unusually enhanced ferromagnetic susceptibility $\chi_0$ roughly scaled by 
$\chi_0 \propto T^{-\zeta}$ and $\chi_0 \propto |H-H_c|^{-\zeta'}$ 
with $\zeta\sim \zeta' \sim 0.6$~\cite{Gegenwart_1} 
contradicting the standard expectation of saturation to a constant.
In accordance, the magnetization shows convex dependence on $H$~\cite{Gegenwart_1}. 
NMR~\cite{Ishida} and ESR~\cite{Sichelschmidt} signals are also consistent.
These non-Fermi-liquid properties are all contradicting the standard theory~\cite{Moriya,Hertz,Millis,Stewart} for the AF QCP and are under extensive debates\cite{Lohneysen}. 

A hint comes from the fact that the first-order transition is observed for YbRh$_{2}$Si$_{2}$
under pressure~\cite{Plessel}.  Actually, the proximity of the first-order transition is common 
in many compounds with unconventional
non-Fermi-liquid properties. 
Our idea is that the proximity of the first-order transition, namely, tricriticality
solves the puzzle because the tricriticality necessarily induces ferromagnetic tendency even at a clear AF transition.
 
At $T\ne 0$, the tricritical point (TCP) where phase
transitions change from continuous to first order
as in Fig. \ref{fig:QTCP_Global_Phase} (b) has been studied in detail~\cite{Sarbach}.
A characteristic feature of TCP is the diverging susceptibility not only at the ordering 
wavenumber $Q$ but also at zero ($\chi_0$)~\cite{Misawa}.  

If quantum fluctuations suppress the temperature of TCP to zero, 
{\it quantum tricritical point} (QTCP) appears 
[see Fig. \ref{fig:QTCP_Global_Phase}(c)].
Then QTCP may alter the criticality of QCP 
as a proximity of the first-order transition. 

Recently TCP has been studied for the {\it itinerant} ferromagnet~\cite{Belitz}
to understand the nature of the global phase diagram of
weak itinerant ferromagnets ZrZn$_2$ and UGe$_2$.
Furthermore, ferromagnetic QTCP has been studied 
for itinerant helical ferromagnet MnSi~\cite{Schmalian}
and nearly ferromagnetic metal Sr$_3$Ru$_2$O$_7$~\cite{Green}.
However, 
these previous studies on the ferromagnetic QTCP do not explain the
unconventional coexistence of the ferromagnetic and
AF fluctuations observed near the AF QCP in YbRh$_2$Si$_2$.

In this letter, we propose that the proximity of the first-order transition
opens a way to solve the puzzles in the AF quantum critical phenomena.  
The proximity of the first-order transition inherently generates diverging ferromagnetic fluctuations concomitantly with the order-parameter (AF) fluctuations. 
The emergence of the concomitance is manifested by the {\it quantum tricriticality}, 
which generates an unexplored non-Fermi liquid.
An unconventional scaling is derived  
by extending the self-consistent renormalization (SCR) theory~\cite{Moriya}
for spin fluctuations.
Our result accounts for the otherwise puzzling properties of YbRh$_{2}$Si$_{2}$,
even when we do not consider the possible valence transitions~\cite{Miyake}
or collapse of $f$-electron itinerancy as in the picture of the 
local quantum criticality~\cite{Coleman}.

To understand the QTCP,
we start from a standard Ginzburg-Landau-Wilson (GLW) expansion
effective action for bozonic spin fields $\varphi_q$ at the wave number $q$~\cite{Moriya,Hertz, Millis}:
\begin{align}
S[ \varphi_{q}]=&\frac{1}{2}\sum_{q}r_{q}| \varphi_{q}|^{2}
+\sum_{q,q^{\prime},q^{\prime\prime}}u(q,q^{\prime},q^{\prime\prime})
(\varphi_{q}\cdot \varphi_{-q^{\prime}}) \notag \\
\times&( \varphi_{q^{\prime\prime}}\cdot \varphi_{q^{\prime}-q-q^{\prime\prime}}) 
+v\sum_{q_1\sim q_5}
( \varphi_{q_1}\cdot \varphi_{-q_2})
( \varphi_{q_3}\cdot \varphi_{-q_4}) \notag \\
\times&( \varphi_{q_5}\cdot \varphi_{q_2+q_4-q_1-q_3-q_5})-H\varphi_{0} \label{Eq:eff},
\end{align}
where $H$ is external magnetic field;  $u(q,q^{\prime},q^{\prime\prime})$ and $v$ are constants,
while $r_{q}$ depends on the magnetic field $H$. 
From eq. (\ref{Eq:eff}),
the free energy $F$ is obtained from 
\begin{equation}
\exp(-F/T)=\int\prod_{q}\mathcal{D}\varphi_{q}\exp(-S[\varphi_{q}]/T).
\end{equation}
Since the QTCP is expressed by fluctuations at both the AF Bragg wavenumber $Q$ and $0$, 
we approximate the free energy as a function of the
order parameter $M^{\dagger}=\langle \varphi_{Q}\rangle$ and the
uniform
magnetization $M=\langle \varphi_{0}\rangle$:
\begin{align}
F_{0}=&\frac{1}{2}\tilde{r}_{Q}{M^{\dagger}}^2+\tilde{u}_{Q}{M^{\dagger}}^4
+v{M^{\dagger}}^6 \notag \\
&+\frac{1}{2}\tilde{r}_{0}M^2
+\tilde{u}_{0}M^4+vM^6-HM, \label{Eq:Free_1}
\end{align}
where $\tilde{r}_{Q}$, $\tilde{u}_{Q}$,
$\tilde{r}_{0}$, $\tilde{u}_{0}$ and $\mathcal{K}$ are defined as
$\tilde{r}_{Q}(T,H)=r_{Q}(H)+12u_{Q}(\mathcal{K}+M^{2}) 
+90v(\mathcal{K}+M^2)^{2}$, 
$\tilde{u}_{Q}(T,H)=u_{Q}+15v(\mathcal{K}+M^2)$,  
$\tilde{r}_{0}(T,H)=r_{0}(H)+12u_{0}\mathcal{K}+90v\mathcal{K}^2$,  
$\tilde{u}_{0}(T,H)=u_{0}+15v\mathcal{K}$, 
and 
\begin{align}
\mathcal{K}=&\sum_{q\neq 0,Q}\langle |\varphi_{q}|^{2}\rangle.\label{Eq:K} 
\end{align} 
Effects of spin fluctuations are included in $\mathcal{K}$
following the SCR theory.
We approximate $u(q,q,Q)$ [$u(q,q,0)$]
and the equivalent coefficients as 
$q$-independent values;
$u(q,q,Q)\simeq u_{Q}$ [$u(q,q,0)\simeq u_{0}$]
for all $q$ [for $q\neq Q$].

We eliminate $M$ in eq. (\ref{Eq:Free_1}) by using the
saddle point condition for $M$, $\partial F_0/\partial M=0$
leading to the relation between $M$ and $M^{\dagger}$ as
\begin{equation}
M=a_{0}+a_{1}{M^{\dagger}}^{2}+a_{2}{M^{\dagger}}^{4}+\cdots, \label{Eq:M_exp}
\end{equation}
where the expansion coefficients $a_{0}\sim a_{2}$ are
determined by substituting eq. (\ref{Eq:M_exp}) into the saddle point condition:
\begin{align}
&\tilde{r}_{0}(T,H)a_{0}+4\tilde{u}_{0}(T,H)a_{0}^{3}+6va_{0}^{5}-H=0, \label{Eq:a_0}\\
&12a_{0}\tilde{u}_{Q}(T,H)+a_{1}R(T,H)=0, \label{Eq:a_1} 
\end{align}
where $R(T,H)=\tilde{r}_{0}(T,H)+12\tilde{u}_{0}(T,H)a_{0}^{2}+30va_{0}^{4}$.
By using eq. (\ref{Eq:M_exp}), we obtain the free energy as
\begin{align}
F_{0}=&\frac{1}{2}\tilde{r}_{Q}(T,H){M^{\dagger}}^{2}+\tilde{u}_{Q}^{\prime}(T,H){M^{\dagger}}^{4} 
+O({M^{\dagger}}^6), \label{Eq:Free_2}
\end{align}
where $\tilde{u}_{Q}^{\prime}(T,H)=\tilde{u}_{Q}(T,H)(1+6a_{0}a_{1})$.
In eq.~(\ref{Eq:Free_2}), continuous phase transitions occur at 
$\tilde{r}_{Q}=0$ when $\tilde{u}_{Q}^{\prime}(T,H)>0$,
while the first-order phase transitions occur when $\tilde{u}_{Q}^{\prime}(T,H)<0$~\cite{Sarbach, Misawa}.
Therefore, the QTCP appears when the conditions $\tilde{r}_{Q}(0,H_{t})=0$
and  $\tilde{u}_{Q}(0,H_{t})=0$ are both satisfied~\cite{comment}, where $H_{t}$
is the critical field at the QTCP.

We now discuss the susceptibilities $\chi_{Q}$ at the AF vector $Q$ and $\chi_{0}$
at $q=0$
in the disordered phase ($M^{\dagger}=0$, $M=a_{0}$) by using eq.~(\ref{Eq:a_0}) and the free energy (\ref{Eq:Free_2}).
From eq. (\ref{Eq:Free_2}),
$\chi_{Q}^{-1}$ is given as
\begin{equation}
\chi_{Q}^{-1}=\frac{\partial^2 F_{0}}{\partial {M^{\dagger}}^2}\Big|_{M^{\dagger}=0}=
\tilde{r}_{Q}(T,H).\label{Eq:chi_Q}
\end{equation}
By differentiating eq.~(\ref{Eq:a_0}) with respect to the magnetic field $H$,
we obtain $\chi_{0}^{-1}$ as
\begin{align}
\chi_{0}^{-1}&\equiv\!\!\Big(\frac{\partial a_{0}}{\partial H}\Big)^{-1}
=\frac{R(T,H)}{1-a_{0}\partial\tilde{r}_{0}/\partial H-4a_{0}^{3}\partial\tilde{u}_{0}/\partial H} \notag \\
&\propto \tilde{u}_{Q}(T,H). \label{Eq:chi_0}
\end{align} 
Here, we used eq.~(\ref{Eq:a_1}), which gives 
$R(T,H)\propto\tilde{u}_{Q}(T,H)$.

Now, the fluctuation-dissipation (FD) theorem~\cite{Kubo} 
\begin{equation}
\sum_{q\neq0, Q}\!\!\langle|\varphi_{q}|^{2}\rangle 
=\frac{2}{\pi}\int_{0}^{\infty}\!\!\!d\omega\Big(\frac{1}{2}+n(\omega)\Big)
\!\!\!\sum_{q\neq0, Q}\!{\rm Im}\chi(q,\omega),
\label{Eq:FD}
\end{equation}
and $n(\omega)\equiv 1/({\rm e}^{\omega/T}-1)$,
combined with eqs. (\ref{Eq:K}), (\ref{Eq:a_0}), (\ref{Eq:chi_Q}), and (\ref{Eq:chi_0})
constitute the self-consistent equations~\cite{Moriya} to determine
$\mathcal{K}$ and $\chi$ in the scheme
of our extended SCR theory.
Using this SCR theory, we now clarify how the susceptibilities
and the magnetization measured from the QTCP
($\chi_{Q}^{-1}$, $\chi_{0}^{-1}$, $\delta a_{0}\equiv a_{0}-a_{0t}$ with
$a_{0t}$ being the value at the QTCP)
are scaled with $\delta H=H-H_{t}$ and $T$ near the QTCP.
The results will be shown in
eqs.~(\ref{Eq:R_chi_Q})-(\ref{Eq:R_d_a0}).   

In the SCR theory,
non-trivial temperature dependence of physical properties
comes from the spin fluctuation term $\mathcal{K}$.
Therefore, we first clarify 
the scaling of $\mathcal{K}$ by using the FD theorem
combined with expansions of $\chi_{0+q}(\omega)$
and $\chi_{Q+q}(\omega)$ in terms of the wavenumber $q$ and
the frequency $\omega$ near the QTCP.
The ordering susceptibility $\chi_{Q+q}(\omega)$
is assumed to follow the conventional Ornstein-Zernike form
($\chi_{Q+q}(\omega)^{-1}\simeq \chi_{Q}^{-1}+A_{Q}q^2-iC_{Q}\omega$) as in
the SCR formalism,
while the uniform part $\chi_{0+q}(\omega)$
turns out {\it not} to follow.
This is because the scaling relation $\chi_{0}^{-1}\propto \chi_{Q}^{-1/2}$
holds near TCP within the GLW theory~\cite{Sarbach}.
As we will see, the self-consistency
among eqs.~(\ref{Eq:K}), (\ref{Eq:a_0}), (\ref{Eq:chi_Q}), (\ref{Eq:chi_0}),
and (\ref{Eq:FD}) requires that this relation still holds for
$q\neq0$.
Therefore, we obtain the relation as
$\chi_{0+q}(0)^{-1}\propto\chi_{Q+q}(0)^{-1/2}\propto(\chi_{Q}^{-1}+A_{Q}q^2)^{1/2} 
\propto(\chi_{0}^{-2}+A_{0}q^2)^{1/2}$.
From the conservation law, $\omega$ dependence of 
$\chi_{0+q}(\omega)^{-1}$ should be given as $\chi_{0+q}(\omega)^{-1}\simeq\chi_{0+q}(0)^{-1}-iC_{0}\omega/q$.
Finally, we obtain 
$\omega$ and $q$ expansions of
$\chi_{0+q}(\omega)^{-1}$ as
$\chi_{0+q}(\omega)^{-1}\simeq(\chi_{0}^{-2}+A_{0}q^2)^{1/2}-iC_{0}\omega/q$.

By substituting the above forms for $\chi_{0+q}(\omega)^{-1}$ [$\chi_{Q+q}(\omega)^{-1}$]
into the FD theorem~(\ref{Eq:FD}),
the contributions from the spin fluctuations near zero [ordering] wavenumber
is given as
$\sum_{q\sim0}\langle \varphi_{q}^{2} \rangle\simeq K_{00}-K_{01}\chi_{0}^{-2}+K_{0T}T^{2}$,
[$\sum_{q\sim Q}\langle \varphi_{q}^{2} \rangle\simeq K_{Q0}-K_{Q1}\chi_{Q}^{-1}+K_{QT}T^{3/2}$]
where $K_{00}$, $K_{01}$, and $K_{0T}$ 
[$K_{Q0}$, $K_{Q1}$, and $K_{QT}$] are constants.
From these relations, in three dimensions, we obtain the scaling of
$\delta\mathcal{K}$ measured from the QTCP as
\begin{equation}
\delta\mathcal{K}\simeq-K_{01}\chi_{0}^{-2}-K_{Q1}\chi_{Q}^{-1}+K_{0T}T^{2}+K_{QT}T^{3/2}.\label{Eq:d_K}
\end{equation}

The singularity of magnetization $a_{0}$ is obtained by solving
eq.~(\ref{Eq:a_0}). Near the QTCP, eq.~(\ref{Eq:a_0}) can be approximated as
$A\delta a_{0}^2+B\delta a_{0}+C=0$,
with $A=12a_{0t}(5va_{0t}^{2}+\tilde{u}_{0})$, $B=\delta\tilde{r}_{0}+12a_{0t}^{2}\delta\tilde{u}_{0}$,
and $C=a_{0t}\delta\tilde{r}_{0}+4a_{0t}^{3}\delta\tilde{u}_{0}-\delta H$,
where $\delta\tilde{r}_{0}=\tilde{r}_{0}(T,H)-\tilde{r}_{0}(0,H_{t})$, and 
$\delta\tilde{u}_{0}=\tilde{u}_{0}(T,H)-\tilde{u}_{0}(0,H_{t})$.
Since both $B$ and $C$ vanish at the QTCP,  
we obtain the asymptotic behavior of $\delta a_{0}$ as
\begin{equation}
\delta a_{0}\simeq (\alpha_{0}\delta H+ \alpha_{1}\delta \mathcal{K})^{1/2}\label{Eq:d_a0},
\end{equation}
where $\alpha_{0}, \alpha_{1}$ are constants. 

By defining $\delta \tilde{r}_{Q}(T,H) \equiv \tilde{r}_{Q}(T,H)-\tilde{r}_{Q}(0,H_{t})$, we get
\begin{align}
\chi_{Q}^{-1}=\delta \tilde{r}_{Q}(T,H)  =&
\delta r_{Q}(H)+90v(\delta\mathcal{K}+\delta{\tilde{a}_{0}})^{2} \label{Eq:chi_Q1},
\end{align}
since both $\tilde{r}_{Q}(0, H_{t})$ and $\tilde{u}_{Q}(0, H_{t})$ 
are zero at the QTCP and terms linear in $\delta\mathcal{K}$ and $\delta \tilde{a}_{0}$ vanish.
Here $\delta r_{Q}$ and $\delta\tilde{a}_{0}$ are
defined as
$\delta r_{Q}=r_{Q}(H)-r_{Q}(H_{t})\simeq r_{QH}\delta H$, 
$\delta \tilde{a}_{0}=a_{0}^2-a_{0t}^2=\delta a_{0}(\delta a_{0}+2a_{0t})$. 

\begin{figure}[h!]
	\begin{center}
		\includegraphics[width=6cm,clip]{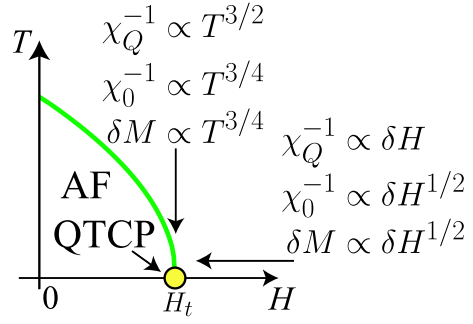}   
	\end{center}
\caption{(color online). Schematic phase diagram around the QTCP
under the magnetic fields $H$.
$\delta M$ denotes the magnetization measured from the critical value at the QTCP.}
\label{fig:Phase_theory}
\end{figure}

From eqs.~(\ref{Eq:d_K}) and (\ref{Eq:d_a0}),
$\delta \mathcal{K}$ is higher order of $\delta a_{0}$ near the QTCP.
Then, from eqs. (\ref{Eq:chi_0}) and (\ref{Eq:chi_Q1}),
the most dominant terms of $\chi_{Q}^{-1}$ and $\chi_{0}^{-1}$
are given as
$\chi_{0}^{-1}\propto \delta a_{0}$, 
$\chi_{Q}^{-1}\simeq r_{QH}\delta H+360va_{0t}^{2}\delta a_{0}^{2}$,
which together with eqs.~(\ref{Eq:d_K}) and (\ref{Eq:d_a0}) leads to
$\delta H$ and $T$ dependence as
\begin{align}
\chi_{Q}^{-1}&\simeq\beta_{Q0}\delta H+\beta_{Q1}T^{3/2},\label{Eq:R_chi_Q}\\
\chi_{0}^{-1}&\simeq(\beta_{00}\delta H+\beta_{01}T^{3/2})^{1/2},\label{Eq:R_chi_0} \\
\delta a_{0}&\simeq(\alpha_{0}^{\prime}\delta H+\alpha_{1}^{\prime}T^{3/2})^{1/2}\label{Eq:R_d_a0}.
\end{align}
Singularities of the uniform susceptibility $\chi_{0}$, 
the ordering susceptibility $\chi_{Q}$,  
and the magnetization $\delta a_{0}$ near the QTCP are summarized in Fig. \ref{fig:Phase_theory}.
We note that the singularity of $\chi_{Q}^{-1}$
given by Green $et$ $al$~\cite{Green} as $T^{8/3}$
is not correct, since they neglect the
$T^{2}$ dependence of the bare second-order coefficient $r_{q}$~\cite{Millis}.

We now examine whether the criticality of the QTCP
is consistent with the experimental results of YbRh$_{2}$Si$_{2}$.
We first emphasize that the puzzling 
critical exponents in YbRh$_{2}$Si$_{2}$ 
described above for $T$ and $H$ dependences of $\chi_0$
and $M$ are well consistent with the quantum tricriticality derived from
eqs. (\ref{Eq:R_chi_0}) and (\ref{Eq:R_d_a0}), while any other theories 
do not reproduce these exponents.

To compare with the experimental results more quantitatively,
we now solve the self-consistent equations numerically,
and obtain 
the uniform susceptibility $\chi_{0}$
and the magnetization $\delta a_{0}$ near the QTCP as follows:
We first approximate the magnetic field dependence of 
$\delta r_{0}(H)\equiv r_{0}(H)-r_{0}(H_{t})$
as $\delta r_{0}(H)\simeq r_{0H}\delta H$.
The contributions from the spin fluctuations, namely $\mathcal{K}$,
can be calculated by setting the four SCR parameters $T_{0A}$,
$T_{00}$, $T_{QA}$, and $T_{Q0}$~\cite{parameter}. 
Furthermore, once the parameters 
$v$, $r_{QH}$, $r_{0H}$, $H_{t}$, and $a_{0t}$ are fixed,
the other parameters ($r_{0}$, $r_{Q}$, $u_{0}$, and $u_{Q}$)
are determined from the conditions $\tilde{r}_{Q}(0,H_{t})=0$,
$\tilde{u}_{Q}(0,H_{t})=0$ and eqs.~(\ref{Eq:a_0}), (\ref{Eq:a_1}).
In this letter, to calculate 
physical properties, we employ a reasonable set of parameters
given in ref.~\citen{parameter_2} as follows:
We estimate $H_{t}$ and $a_{0t}$ directly from experiments and also choose
conventional SCR parameters ($T_{0A}, T_{00}, T_{QA}, T_{Q0}$) 
within the order of 10-100K. This range of SCR parameters is typical 
in heavy fermion compounds~\cite{Moriya}.
In contrast to these, for the other non-primary parameters ($r_{QH}$, $r_{0H}$, and $v$),
we do not find any constraint from physical requirement.
Therefore, we have freely tuned these parameters to reproduce
the experimental results quantitatively.
However, the critical exponents do not change 
even if we have chosen these parameters arbitrarily.
Microscopic derivation of these phenomelogical
parameters is left for future studies.

\begin{figure}[h!]
	\begin{center}
		\includegraphics[width=8.5cm,clip]{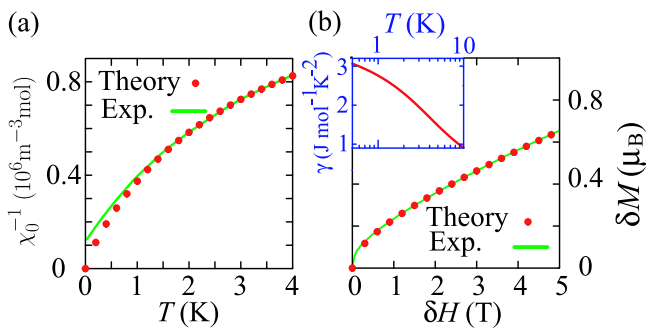}   
	\end{center}
\caption{(color online). (a) Experimental $\chi_{0}^{-1}$
for YbRh$_{2}$(Si$_{0.95}$Ge$_{0.05}$)$_{2}$
at $H=0.03$(T) reported in ref.~\citen{Gegenwart_1}
compared with the present SCR theory. Green line (red circle) represents the
experimental (theoretical) $\chi_{0}^{-1}$. Theoretical $\chi_{0}^{-1}$ is
calculated just on the QTCP ($H=H_{t}$).
(b) Experimental magnetization curve
for YbRh$_{2}$(Si$_{0.95}$Ge$_{0.05}$)$_{2}$
at $T=0.09$(K) reported in ref.~\citen{Gegenwart_1}
compared with the present theory. 
Green line (red circle) represents the
experimental (theoretical) magnetization curve. 
$\delta M$ ($\delta H$) represents the magnetization (magnetic field)
measured from the critical value.
We estimate the experimental critical magnetic field $H_{c}$ (magnetization $M_{c}$)
as 0.027(T) (0.004 ($\mu_{\rm B}$)).
The inset shows the numerical result of Sommerfeld coefficient of the specific heat $\gamma$
just on the QTCP.
}
\label{fig:Fig3}
\end{figure}

In Fig.~\ref{fig:Fig3} (a),
the numerical result of the 
temperature dependence of $\chi_{0}^{-1}$
just on the QTCP is compared with the experimental $\chi_{0}^{-1}$
reported in ref.~\citen{Gegenwart_1}.
Although we obtain the critical exponent 
$\zeta=0.75$ $(\chi_{0}^{-1}\propto T^{\zeta})$ for $T\rightarrow 0$,
the numerical result shows that $\chi_0^{-1}$ is roughly scaled by $T^{0.6}$ 
at higher temperatures ($T>$1.0K). 
We emphasize that the puzzling convex behavior
of $\chi_{0}^{-1}$ for YbRh$_{2}$(Si$_{0.95}$Ge$_{0.05}$)$_{2}$
near the QCP can not be accounted for by the
conventional quantum criticality, because the critical exponent
$\zeta$ is always larger than one for the conventional 
quantum critical point~\cite{Moriya,Hertz,Millis}.  
The nonzero offset of experimental $\chi_{0}^{-1}$
at $T=0$ indicates that the QCP in 
YbRh$_{2}$(Si$_{0.95}$Ge$_{0.05}$)$_{2}$ 
exists slightly away from the QTCP.
It is an intriguing experimental challenge
to determine the precise location of the QTCP
by tuning the pressure and the magnetic field.

In Fig.~\ref{fig:Fig3} (b),
the numerical result of magnetization curve 
at $T=0$ is compared with the experimental result
reported in ref.~\citen{Gegenwart_1}.
By using the same parameters, 
it is noteworthy that
not only the uniform susceptibility 
but also the magnetization is consistent with the
experimental result quantitatively
in the relevant parameter region.
Because the experimentally observed QCP is slightly away from the
QTCP, small deviations are seen at low temperatures and low magnetic fields
($T<1.0$K and $\delta H < 0.5$T). 
We note that this singularity of the magnetization 
[$\delta M\propto \delta H^{1/2}$ as seen from eq.~(\ref{Eq:R_d_a0}) and Fig.~\ref{fig:Fig3} (b)]
is qualitatively different from that of the conventional
metamagnetic transitions. It has been proposed that they belong to the Ising 
universality class~\cite{QCEP}.
The critical exponent $\delta$ ($\delta M\propto \delta H^{1/\delta}$) of the Ising universality is always larger than three at any dimensions.
The present critical exponent $\delta=2$ makes a sharp contrast
to such conventional critical exponents $\delta\ge 3$.

We now discuss the singularity of the
Sommerfeld coefficient of the specific heat $\gamma$.
In the inset of Fig.~\ref{fig:Fig3} (b),
the numerical result of $\gamma$
just on the QTCP is shown.
At high temperatures ($T>1.0$K),
both the singularity ($\gamma \propto -\log{T}$)
and the amplitude are consistent with 
those of the experimental result~\cite{Custers}.
However, at low temperatures ($T<1.0$K), within this SCR theory,
$\gamma$ near the AF QTCP approaches that of the conventional
AF QCP ($\gamma \propto {\rm const.}-T^{1/2}$),
while experimentally, 
power-law-like behavior is observed for $T<0.3K$~\cite{Custers}.
This discrepancy may be solved by considering either
the fact that the N$\acute{{\rm e}}$el temperature is actually
nonzero or effects of valence fluctuations~\cite{Miyake},
while the criticality of magnetic properties analyzed 
here should remain unchanged.

Finally, we discuss a different
scenario for the QCP in YbRh$_{2}$Si$_{2}$ proposed by Coleman $et$ $al$~\cite{Coleman}.
They claim that a breakdown of composite heavy fermion
(namely, all $f$ electrons decouple from the Fermi surface) occurs at the QCP
and no heavy electron exists in the ordered phase any more.
This scenario is inconsistent with
a large Sommerfeld coefficient of 
the specific heat $\gamma$
even in the ordered phase~\cite{Custers}.
While a large change of Hall constant
in YbRh$_{2}$Si$_{2}$~\cite{Paschen} 
was suggested to support
their scenario,
Norman~\cite{Norman} has pointed out that small changes of the
$f$ electron occupation are sufficient to reproduce the experimental
result by calculating the band structures of YbRh$_{2}$Si$_{2}$. 
Steep change in the Hall coefficient is then naturally understood under the proximity of the 
first-order transition.

In summary, a non-Fermi liquid 
different from that obtained from the conventional QCP is shown to
emerge when the proximity of the first-order transition is involved through the 
QTCP.  The unconventional criticality thus obtained by the extension of the SCR theory
solves the puzzles in the experimental results of YbRh$_{2}$Si$_{2}$.
It is intriguing to examine whether this proximity also plays roles in other unconventional
non-Fermi liquids.

\acknowledgements{This work is supported by Grant-in-Aid for Scientific Research  
under the grant numbers 17071003 and 16076212 from 
MEXT, Japan.
TM is supported by the Japan Society for the Promotion of Science.}


\begin{thebibliography}{999}
\bibitem{comment}
We assume that $1+6a_{0}a_{1}$ does not
vanish at the QTCP.
\bibitem{parameter}
We define as $T_{0A}=(A_{0}q_{B}^{2}/4)^{1/2}$, $T_{00}=q_{B}T_{0A}/\pi C_{0}$
$T_{QA}=A_{Q}q_{B}^{2}/2$, and $T_{Q0}=T_{QA}/\pi C_{Q}$, where
$q_B$ is the effective zone-boundary vector.
\bibitem{parameter_2}
From the experimental result,
we choose the parameters for the QTCP as $H_{t}=0.08$(T) and $a_{0t}=0.2$($\mu_{\rm B}$).
The other parameters are chosen as
$T_{0A}=20$(K), $T_{00}=10$(K), $T_{QA}$=270(K), $T_{Q0}$=5(K),
$v=21$(K), $r_{0H}=5$, and $r_{QH}=180$.
\end{thebibliography}
\end{document}